\documentclass[aps,pra,twocolumn,groupedaddress,reprint]{revtex4}
\usepackage{graphicx}
\usepackage{setspace}
\usepackage{amsmath}
\usepackage{amssymb}
\usepackage{color}
\usepackage{array}
\usepackage{subfigure}
\usepackage{hyperref}
\usepackage{float}
\usepackage{lipsum}
\usepackage{amsfonts}
\usepackage{mathrsfs}
\usepackage{dcolumn}
\usepackage{bm}
\usepackage{bbm}
\usepackage{graphicx}
\usepackage{epstopdf}

\begin{document}
\title{Experimental study of quantum random number generator based on two independent lasers}
\author{Shi-Hai Sun$^1$}
\altaffiliation{shsun@nudt.edu.cn}
\author{Feihu Xu$^{2,3}$}
\altaffiliation{fhxu@mit.edu}

\affiliation{$^1$ College of Science, National University of Defense Technology, Changsha 410073, P.R.China\\
$^2$ Research Laboratory of Electronics, Massachusetts Institute of Technology, 77 Massachusetts Avenue, Cambridge, Massachusetts 02139, USA\\
$^3$ Hefei National Laboratory for Physical Sciences at Microscale, Shanghai Branch, University of Science and Technology of China, Hefei, Anhui 230026, China}

\date{\today}
\begin{abstract}
Quantum random number generator (QRNG) can produce true randomness by utilizing the inherent probabilistic nature of quantum mechanics. Recently, the spontaneous-emission quantum phase noise of the laser has been widely deployed for QRNG, due to its high rate, low cost and the feasibility of chip-scale integration. Here, we perform a comprehensive experimental study of phase-noise based QRNG with two independent lasers, each of which operates in either continuous-wave (CW) or pulsed mode. We implement QRNGs by operating the two lasers in three configurations, namely CW+CW, CW+pulsed and pulsed+pulsed, and demonstrate their tradeoffs, strengths and weaknesses.
\end{abstract}


\maketitle
\textbf{\emph{Introduction-}} True randomness plays an important role in widespread applications, and it is generally believed to be impossible using only classical process. Quantum random number generator (QRNG), however, can generate true and unpredictable random numbers by exploiting the inherent randomness of quantum mechanics \cite{Ma-Yuan16,Herrero-Collantes17}. During the past decade, QRNG has been implemented that is based on different types of quantum phenomenons including single-photon detection \cite{Rarity94,Jennewein00,Dynes08,Wayne09,Furst10,L15,Xu16}, vacuum fluctuations \cite{Gabriel10,Shen-Tian10,Haw15,Marangon17}, phase noise \cite{Guo-Tang10,Bing-Chi10,Jofre11,Xu-Qi12,ma13,Yuan14,Nie15,Abellan15} of amplified spontaneous emission, and quantum non-locality \cite{Pironio10}.

Among these implementations, the quantum (spontaneous-emission) phase noise of a laser has the advantages of high rate and low cost, which has attracted a lot of scientific attention \cite{Guo-Tang10,Bing-Chi10,Jofre11,Xu-Qi12,ma13,Yuan14,Nie15,Abellan15}. In previous QRNGs, an unbalanced interferometer was generally employed to measure the quantum phase noise. However, such an implementation has two practical drawbacks: first, it requires the phase stability of the interferometer; second, the large footprint of the interferometer makes it unsuitable for chip integration. Very recently, an important scheme which relies upon the interference between two independent lasers -- a continuous-wave (CW) laser and a pulsed laser -- has been proposed and demonstrated to solve these drawbacks \cite{Abellan16,Zhou17,Qi16}, although the quantum randomness and the classical noise (e.g., detector's electrical noise) were not rigorously quantified. These recent works \cite{Abellan16,Zhou17,Qi16} demonstrate the large potential of practical QRNGs with independent lasers as the quantum entropy source. Besides QRNG, the interference between two independent lasers is also valuable to the field of quantum cryptography \cite{Curty10,Lo12}.

In this paper, we present an extensive experimental study of QRNG based on two independent lasers. We operate the two lasers in three configurations, i.e., CW+CW, CW+pulsed, and pulsed+pulsed, and analyze their strengths and weaknesses. By using off-the-shelf fiber-optical components, we demonstrate the maximum random number generation rates under those operating configurations. Moreover, the conditional min-entropy is estimated given the uniform distribution of the practical quantum phase signal. And both the classical electrical noise and the intensity fluctuation noise of the two lasers are also taken into account. Thus, our work provides an important step towards a fast, low-cost, robust QRNG.

\begin{figure}
\scalebox{1}{\includegraphics[width=\columnwidth]{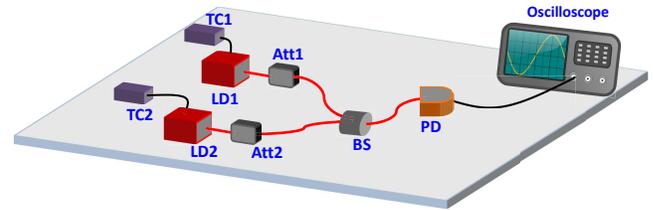}}
\caption{\label{fig:scheme}(Color online) Experimental setup for QRNG. Two independent lasers (LD1 and LD2) interfere at a beam splitter (BS) whose output is detected by a photodetector (PD) followed by an oscilloscope. The output voltage of the photodetector is AC-coupled to the oscilloscope whose sampling rate is determined by the confirguration (see main text). Each analog sample is converted to 8 digital bits by an 8-bit analog-to-digital converter (ADC).}
\end{figure}

\textbf{\emph{Experimental setup-}} The experimental setup is shown in Fig.\ref{fig:scheme}. Two independent distributed feedback (DFB) lasers (LD1 and LD2), followed by two optical attenuators (Att), interfered at a 50:50 beam splitter (BS). The interference signal was detected by a photodetector (PD) with bandwidth 1 GHz (Newport 1611), whose output was sampled by a high-speed oscilloscope (Agilent, DSO9104A). To interfere properly, LD1 and LD2 should be indistinguishable in the dimension of spatial mode, polarization, spectrum, and arrival time. Because single-mode polarization maintaining fibers were used to connect all the optical devices (LDs, Att, BS), the spatial modes and polarizations of LD1 and LD2 were matched automatically. The spectra of the two lasers were controlled by two independent temperature controllers (TC). The 3-dB widths of LD1 and LD2 spectra are both $\sim$17 pm. By carefully adjusting the temperature of LD1 and LD2 independently, the difference of the center wavelength between LD1 and LD2 was made to be much smaller than the lasers' 3-dB linewidths. In our experiment, the two lasers were operated in three cases: (I) CW+CW; (II) CW+pulsed; (III) pulsed+pulsed. Note that, the arrival-time mismatch between LD1 and LD2 affects the interference of case (III) only.

\textbf{\emph{Model-}} By controlling the temperature of LD1 and LD2, the center wavelength of LD1 is brought close to that of LD2. After removing the DC background, the output from the photodetector can be written as
\begin{equation}\label{Eq:voltage}
V(t) \propto E_1E_2\cos[\Delta\omega_0 t+\theta_1(t)-\theta_2(t)],
\end{equation}
where $\Delta\omega_0=\omega_1^0-\omega_2^0$ is the beating frequency, $E_j$ and $\omega_j^0$ ($j=1,2$) are the amplitude and center frequency of j-th laser, $\theta_1(t)$ and $\theta_2(t)$ are the phase of LD1 and LD2, which mainly originates from the quantum phase noise due to spontaneous emission photons~\cite{Yariv07}.

\begin{figure}
\scalebox{1}{\includegraphics[width=\columnwidth]{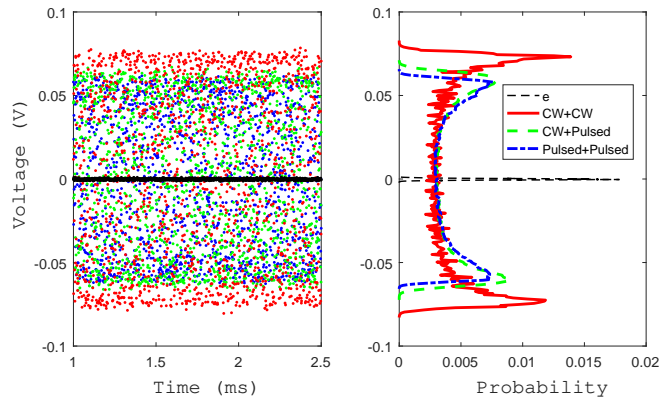}}
\caption{\label{fig:signal_noise}(Color online) The measured output voltage (left) and the probability density function (right) for the electrical (e) noise (black-dashed) and the total phase noise, CW+CW (red-solid), CW+Pulsed (green-dashed) and Pulsed+Pulsed (blue-dash-dotted). The classical noise includes the classical phase noise coming from the interferometer, the intensity fluctuation noise of two lasers, and the classical electrical noise of photodetector and oscilloscope. But as the analysis in the Appendix, the classical phase noise of interferometer is ignored here, and a detailed figure of the second and third classical noises is shown in the appendix. In all measurements, the sampling rate of the oscilloscope is 20 GHz, and the recorded sample size is $10^8$. The maximal amplitude of phase noises are 82.9 mV, 70.8 mV, and 65.7 mV for CW+CW, CW+pulsed and pulsed+pulsed, respectively.}
\end{figure}

This quantum phase noise constitutes our quantum signal. Besides the quantum signal, the output of the PD also contains classical noise, which includes the classical phase noise of the interferometer, the intensity fluctuation noise of the two lasers, and the electrical noise of the detection devices (PD and oscilloscope). Fig.\ref{fig:signal_noise} and Appendix show the measured classical noise. In QRNG, one has to quantify the amount of quantum signal and classical noise in order to extract the genuine quantum randomness in post-processing \cite{Gabriel10,ma13,Haw15,Mitchell15}. We estimate the conditional min-entropy for all of the three cases (CW+CW, CW+Pulse, and Pulse+Pulse) given the classical noise taken into account. These details are shown in the Appendix.

\emph{Case I: CW + CW.} In case I, both LD1 and LD2 were operated in CW mode. By carefully controlling the temperature via TC1 and TC2, the difference of the center wavelength between LD1 and LD2 can be made smaller than the 3-dB linewidths of the two lasers. Then $\theta_j(t)$ ($j=1,2$) can be treated as Gaussian white noise. The variance of $\Delta\theta(t)$ is given by~\cite{Bing-Chi10,Yariv07}
\begin{equation}\label{Eq:variance_dtheta}
\langle[\Delta\theta(t)]^2\rangle=2T_s(\frac{1}{\tau_{c_1}}+\frac{1}{\tau_{c_2}}),
\end{equation}
where $T_s$ is the sampling period and $\tau_{c_j}$ is the coherence time of the j-th laser.

Here, we remark that only the phase noise coming from the spontaneous emission is considered as quantum phase noise in Eq.\ref{Eq:variance_dtheta}. Strictly speaking, the classical phase noise, such as the classical phase noise of the interferometer (due to the mechanical vibration and thermal effect), the intensity fluctuation of light (due to the fluctuation of the driving current of laser and the fluctuation of transmittance of the optical setups) and the classical electrical noise of the measured devices (including the photodetector and the oscilloscope) will contribute a small portion of the randomness output. In the Appendix A, we show how to remove such classical phase noise.

\begin{figure*}[htb]
\centering \subfigure[] {\includegraphics [width=7cm]
{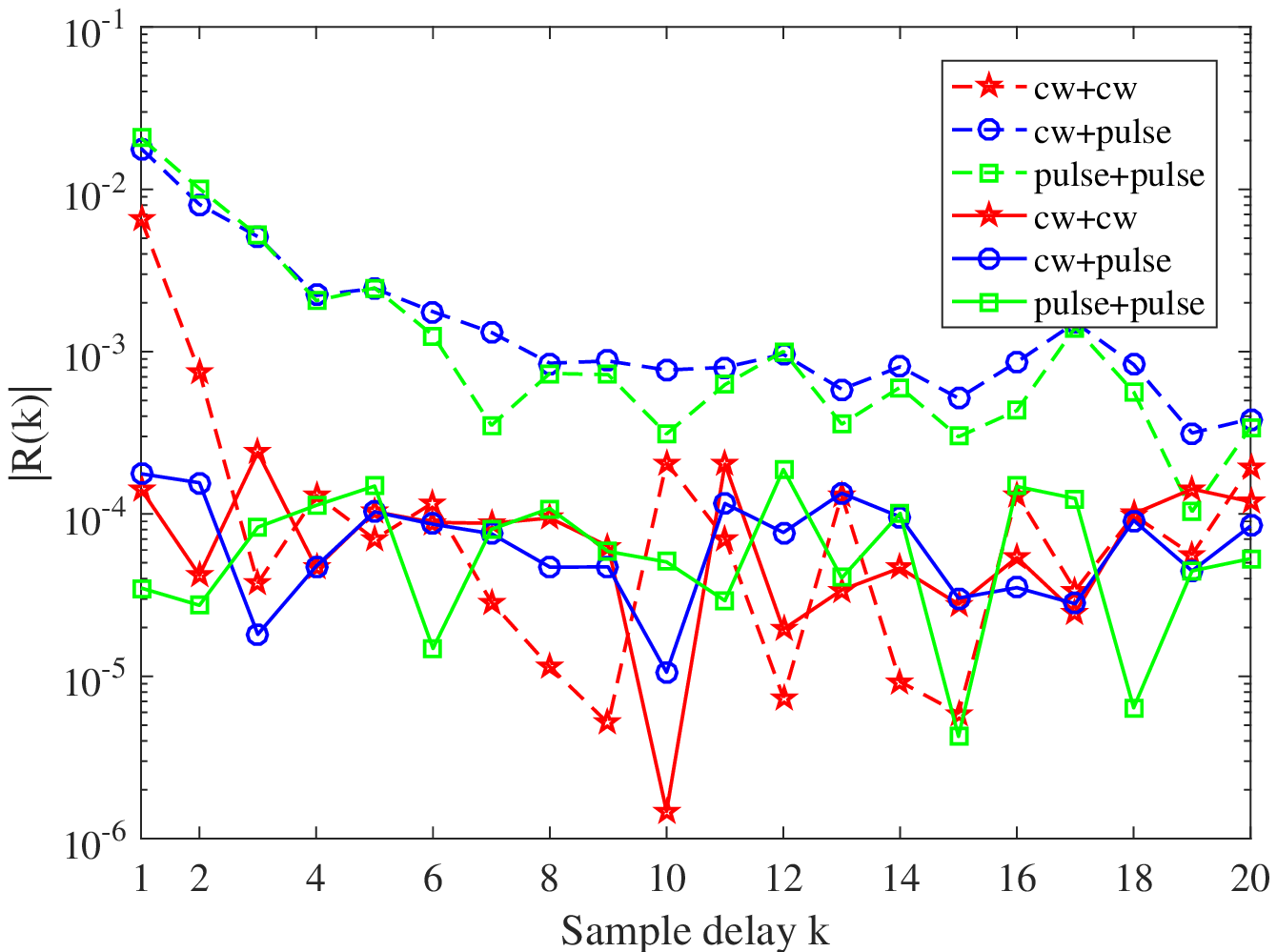} \label{fig:auto_final}} \qquad \subfigure[]
{\includegraphics [width=6.65cm] {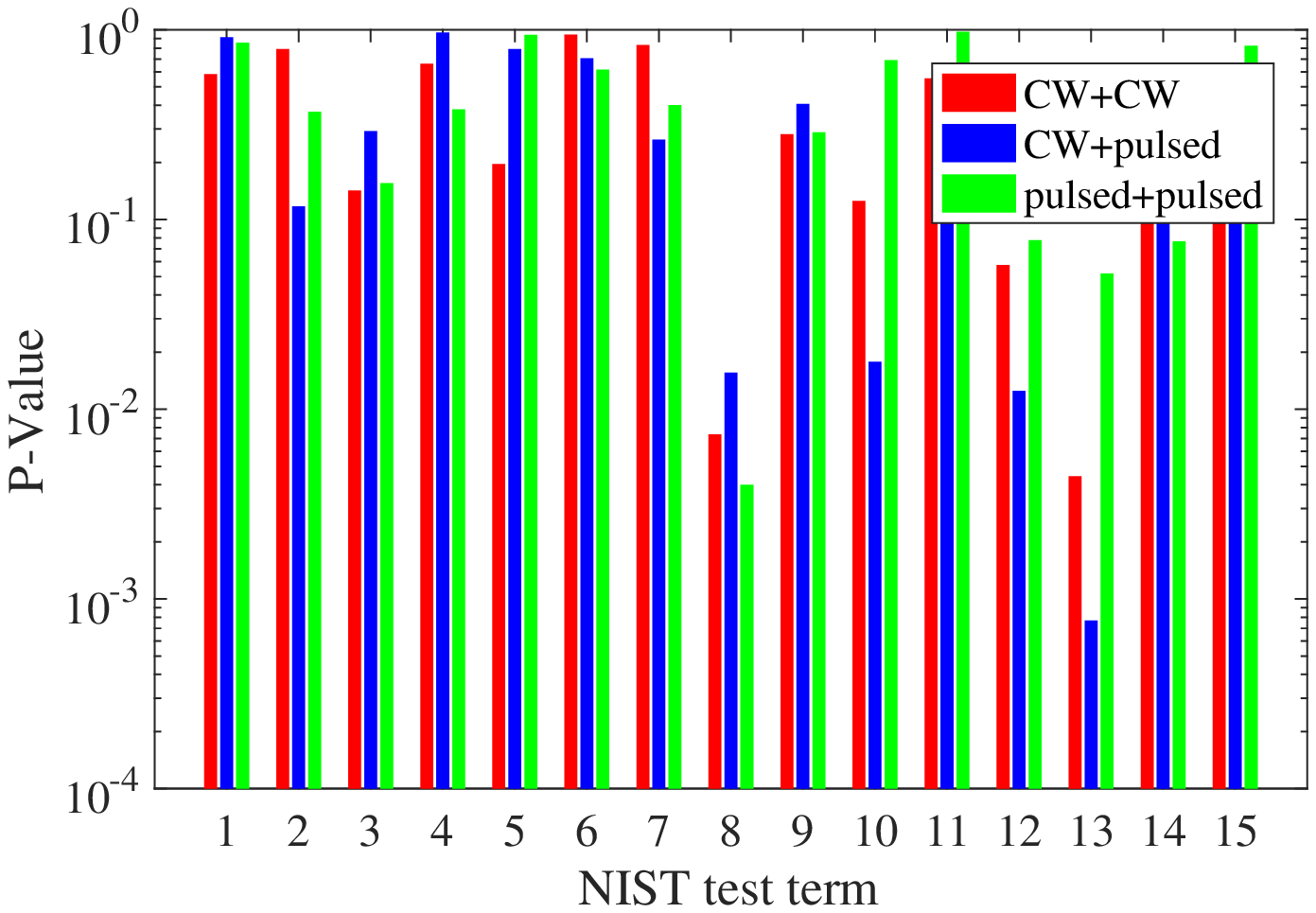}
\label{fig:nisttest}}
\caption{ (Color online) (a) The autocorrelation coefficient of raw data (dotted curves) and extracted data (solid curves) for case I (red star), case II (blue circle), and case III (green square). The autocorrelation coefficient is calculated with the length of data $10^8$ and statistical significance $\alpha=0.01$. And the calculated P-value of the autocorrelation coefficient is given in the Fig.\ref{fig:auto_pvalue} of Appendix, all of them are larger than $\alpha=0.01$. In order to show the autocorrelation coefficient clearly, the absolute value $|R(k)|$ is plotted in the figure. The autocorrelation coefficient of extracted data is rather low (normally below 0.001), which means high quality random number is obtained after the post-processing. (b) NIST test results for 1 Gbits extracted data in each case. The labels in X-axis represent the 15 NIST test terms. In each test term, the bar values from left to right represent the P-values for the three cases~\cite{NIST}. The extracted data successfully passes all NIST tests in all three cases.}
\end{figure*}

According to Eq.\ref{Eq:variance_dtheta}, although the spontaneous emission phase follows Gaussian whit noise, if $T_s\gg \tau_{c_j}$, the variance of the $\Delta\theta(t)$ is $\langle[\Delta\theta(t)]^2\rangle \gg 1$. Then the phase of the two adjacent sampling points is close to be independent. In other words, there is no obvious phase correlation between the two adjacent sampling points. Thus, the phase noise $\Delta\theta$ is close to be a uniform distribution within $(-\pi,\pi]$. In our experiment, the 3dB linewidth of LD1 and LD2 are 0.0177nm and 0.0172nm, respectively. Thus, the coherent time is 459.78ps for LD1 and 473.14ps for LD2. According to Eq.\ref{Eq:variance_dtheta}, the sampling period should satisfy that $T_s > \tau_{c_1}\tau_{c_2}/2(\tau_{c_1}+\tau_{c_2})\approx 116.6ps$. It seams that the sampling rate can be reached to about 8.6GHz.

However, here we must note that, different from the previous works that generate quantum random number with one cw laser \cite{Bing-Chi10}, the sampling rate in the CW+CW case is limited not only by the coherent time of the two lasers, but also the beating frequency of the two lasers. In fact, it is challenging to perfectly match the frequency of the two independent lasers in the CW+CW case, which means $\Delta \omega_0 \neq 0$ (in the one laser case, the frequencies of lights, which interfere at the BS, are perfectly matched). Thus, the measured output voltage signal, $V(t)$, is a sine wave with a phase jitter in our experiment. The center frequency of the since wave comes from the beating frequency $\Delta\omega_0$, and the phase jitter comes from the random phase of the two lasers $\theta_1(t)-\theta_2(t)$. In our experiments, the center frequency of the measured output voltage, $V(t)$, is 278.7MHz with a standard deviation 30.2MHz (see Fig.\ref{fig:cw2_fv} in the Appendix for detail).

Obviously, the higher the sampling rate is, the higher the correlation between the adjacent sampling point is. Thus, in order to evaluate the sampling rate of the experiment, we calculate the autocorrelation coefficient of the output signal at various sampling frequencies, $f_s=1/T_s$. The autocorrelation coefficient $R$ of a sequence $X$ is defined as
\begin{equation}\label{Eq:auto_cor}
R(k)=\frac{E[(x_i-\mu)(x_{i+k}-\mu)]}{\sigma^2},
\end{equation}
where $E$ is the expected value operator, $k$ is the sample delay, $\mu$ and $\sigma$ are the mean and the standard deviation of $X$. In Fig.\ref{fig:autoco_diff_fs}, we show the measured autocorrelation coefficient of the output signal with different $f_s$. In the case that $k$=$1$, the autocorrelation coefficients are 0.2319, -0.1497, and 0.008 for $f_s$=100 MHz, 50 MHz, and 20 MHz.

Thus, in order to remove the classical correlation of sampling point as many as possible, we select the sampling frequency at 20 MHz in our experiment, i.e., $T_s=50$ ns. Then the variance of $\Delta\theta(t)$ is $\langle[\Delta\theta(t)]^2\rangle \approx 428.85 \gg 1$, which means the sampling period is much larger than the coherent time of the lasers. Hence, we can treat $\Delta\theta$ as a uniform distribution within $(-\pi,\pi]$, i.e., the probability density function (PDF) of the quantum signal can be written as
\begin{equation}\label{Eq:2cw_fq}
f_{Q}(q)=\begin{cases}  \frac{1}{\pi\sqrt{A^2-q^2}},& -A < q< A,\\ 0, & \text{else}, \end{cases},
\end{equation}
where $A$ is the maximum amplitude of the quantum signal. The red-solid curve in Fig.~\ref{fig:signal_noise} shows the PDF of the measured output voltage of the PD. It follows an arcsine distribution. The maximum (and minimum) output voltage of the PD is 89.2mV (and -89.2 mV), which is much larger than the amplitude of the classical electrical noise and the intensity fluctuation noise of lasers (both of them are smaller than 2.5mV, see Fig.4 for details). For simplicity, we choose $A=82.9$. Plugging the PDF (Eq.\eqref{Eq:2cw_fq}) into the min-entropy model (see Appendix), we estimate the quantum min-entropy as 4.47 bits per 8-bit sample.

To extract quantum randomness, randomness extractor is always required. In practice, two typical extractors, the Toeplitz-hashing extractor \cite{Wegman81} and the Trevisan's extractor \cite{Trevisan01}, are often used. Both of them have been proven to be information-theoretically secure and taken finite size effects into account. In this paper, as a proof-of-principle demonstration, only the Toeplitz-hashing extractor is used. And a more rigorous discussion of randomness extractor can be found in Ref.\cite{Ma-Yuan16,Herrero-Collantes17}.

The Toeplitz-hashing randomness extractor extracts random bit-string \emph{m} by multiplying the raw sequence \emph{n} with the Toeplitz matrix. In our experiment, we set the size of the Toeplitz matrix is $n=4096$ and $m=2048$ for the simplicity of software implementation, i.e., 4 bits are produced for each sample (8-bit ADC). Finally, the QRNG rate is $20 \times 4=80$ Mbps. After the extraction, we tested the autocorrelation coefficient (Fig.~\ref{fig:auto_final}) and ran the statistic test suite of NIST with successful passes in all tests (Fig.~\ref{fig:nisttest}).

\emph{Case II: CW + PULSED.} In the second case, LD1 operates in CW mode, while LD2 operates in pulsed mode. This scheme has been recently demonstrated in~\cite{Abellan16,Zhou17}. In our experiment, LD2's repetition rate is 500 MHz, and its 3-dB width is 433.2 ps with the standard deviation 27.0 ps. Since the phase of each pulse comes from fresh spontaneous emission photons, $\Delta\theta(t)$ follows a uniform distribution within $(-\pi,\pi]$. This means that the PDF of the quantum signal is the same as Eq.~\eqref{Eq:2cw_fq}.

In our experiment, the interference signal is detected by the PD with bandwidth 1 GHz, followed by an oscilloscope at sampling rate 10 GHz. The green-dashed curve in Fig.~\ref{fig:signal_noise} shows the PDF of the measured output voltage.
The maximum (and minimum) output voltage of the PD is 79.2 mV (and -62.4 mV). Thus, we set the offset of the 8-bit ADC at 8.4 mV, which means that the output number of the ADC is 0 for input voltage -62.4 mV, and 255 for input voltage 79.2 mV. In other words, the parameter $A=70.8$ mV in Eq.~\eqref{Eq:2cw_fq}. Note that although the amplitude of the phase noise could be further increased by increasing the power of the CW laser, this would result in almost no change in the conditional quantum min-entropy. With the same method as case I, we get the min-entropy of quantum randomness as 4.45 bits, and the final QRNG rate is thus $500 \times 4=2$ Gbps. The autocorrelation coefficient and the NIST test results are shown in Fig.\ref{fig:auto_final} and \ref{fig:nisttest}.

\emph{Case III: PULSED + PULSED.} In this case, both LD1 and LD2 operate in pulsed mode with a repetition rate of 500 MHz. The trigger of the two lasers is generated from an arbitrary waveform generator (Tektronix AWG7092C). To guarantee that the two pulses from the two lasers can properly interfere at the BS, the arrival time of the two pulses should be indistinguishable. This was achieved by delaying the triggers for LD1 and LD2 with 1 ps time resolution. In the experiment, the 3-dB temporal widths of the pulses from LD1 and LD2 are 530.9 ps with the standard deviation 19.1 ps and 563.0 ps with the standard deviation 18.9 ps, respectively. By fine-tuning the trigger delay, we can control the overlap of the two pulses at a resolution that is much smaller than the temporal width of the pulses.

Since the phase of each pulse comes from different spontaneous emission photons, the phase follows a uniform distribution. The output voltage also follows an arcsine distribution (see blue dashed-dot curve of  Fig.~\ref{fig:signal_noise}). The measured maximum (and minimum) voltage is 111.6 mV (and -19.8 mV). Then by setting the offset of the ADC at 45.9 mV, we get the parameter $A=65.7$ in Eq.~\eqref{Eq:2cw_fq}. Performing the same method as case I, the estimated min-entropy is 4.43bits, and final random number generation rate is $500 \times 4=2$ Gbps after the post-processing. The autocorrelation coefficient and NIST test results are shown in Fig.\ref{fig:auto_final} and \ref{fig:nisttest}.

\textbf{\emph{Discussion-}} We discuss the tradeoffs of the three cases for practical QRNG with independent lasers. First, among three cases, case I -- CW+CW -- is the easiest to implement, because it does not require any high-speed electronics to modulate optical pulses. And yet it is difficult to match the frequencies of LD1 and LD2, so the beating frequency limits the maximum QRNG rate. In our experiment, the generation rate is limited to tens of Mbps. Second, case II and case III achieve the same generation rate of 2 Gbps. This is because the generation rate primarily depends on the repetition rate of the pulsed laser and the precision of the ADC. Generally speaking, the repetition rate of the pulsed laser can be increased to a few GHz and a generation rate of tens of Gbps is achievable for both case II and case III with current technology. This rate is much higher than case I. Third, case III requires the precise matching of the pulse arrival times for LD1 and LD2. This might be a practical challenge for ultrahigh speed implementations when the repetition rate of the pulsed laser reaches tens of GHz. Therefore, we conclude that case II, cw+pulse, may be the best choice for high-speed implementation of QRNG based on quantum phase noise (as demonstrated recently in \cite{Abellan16,Zhou17}), while case I is suitable for simple and low-cost applications that may require slow rate QRNG only. Notice that high-speed implementation of case I is still possible by changing the scheme to a broadband source and a homodyne detection~\cite{Qi16}.

Overall, we have experimentally demonstrated QRNGs based on two independent lasers. We operated the two lasers in three cases (CW+CW, CW+pulsed and pulsed+pulsed), experimentally studied the properties and tradeoffs for QRNG in each case, and generated truly random numbers at rates of 80 Mbps, 2 Gbps and 2 Gbps, respectively. Our work demonstrates the great potential of quantum phase-noise based QRNG using two independent lasers.

\textbf{\emph{Acknowledgments-}} This work was supported by National Natural Science Foundation of China Grants No. 11674397 and No. 61771443, and Canadian NSERC PDF. The authors thank De-Feng Gu, Connor Henley and Bing Qi for helpful discussions.

\appendix
\section{The estimation of min-entropy}
Taking the classical noise into account, the measured total signal $M$ is $M=Q+N$. Here $Q$ is the quantum signal with probability density function (PDF) $f_{Q}(q)$, and $N$ is the classical noise signal with PDF $f_N(n)$ . Then the PDF of $M$ is the convolution of $f_{Q}(q)$ and $f_N(n)$, which is given by
\begin{equation}
f_M(m)=\int_{-\infty}^\infty f_Q(m-n)f_N(n)dn.
\end{equation}
To evaluate the min-entropy of genuine quantum randomness, we need to estimate the conditional PDF of $M$ given the classical noise $N$. With the same method as Ref.\cite{Haw15}, we can get the conditional cumulative distribution function (CDF), $F_{M|N}(m|n)$, which can be written as
\begin{equation}
\begin{split}
F_{M|N}(m|n)&=P\{M\leq m|N=n\}=\frac{P\{M\leq m, N=n\}}{P\{N=n\}}\\
&=\frac{P\{q\leq m-n, N=n\}}{P\{N=e\}}\\
&=\frac{P\{Q\leq m-n\}\cdot P\{N=n\}}{P\{N=e\}}\\
&=P\{q\leq m-n\}\\
&=F_Q(m-n),
\end{split}
\end{equation}
where $F_Q(q)$ is the CDF of the quantum signal $Q$. Hence it is easy to get the conditional PDF, which is
\begin{equation}\label{Eq:Pme}
f_{M|N}(m|n)=f_Q(m-n),
\end{equation}
where $f_Q(m-n)$ denotes the PDF of the quantum signal $Q$.

By sampling the output voltage of the photodetector with a k-bit ADC, the discretized conditional probability of the measured signal given the classical noise can be written as
\begin{equation}\label{Eq:con_pdf}
P_{M_{dis}|N}(m_i|n)=\int_{V_l+i \delta}^{V_l+(i+1)\delta} f_{M|N}(m|n)dm,
\end{equation}
where $\delta=(V_u-V_l)/2^k$ and \emph{k} is the precision of the ADC. $V_l$ and $V_u$ are the lower and upper bound of the sample range $[V_l,V_u]$. $i=0,1,......,2^k-1$ is the output digital number of ADC. Therefore, the worst-case min-entropy conditioned on classical noise $E$ is given by \cite{Haw15}
\begin{equation}\label{Eq:min_entropy}
H_{min}(M_{dis|N})=-\log_2[\max_{n\in[n_{min},n_{max}]}\max_{m_i\in M_{dis}} P_{M_{dis}|N}(m_i|n)].
\end{equation}
Here the \emph{worst-case} means that, from the adversary's perspective, the classical noise is fully known and controlled by her with arbitrary precision. Therefore, if we know the bound of the classical noise, $n_{min}$ and $n_{max}$, we can estimate the min-entropy with Eq.\ref{Eq:min_entropy}, and then distill true quantum random number by performing the post-processing.

\begin{figure}
\scalebox{1}{\includegraphics[width=\columnwidth]{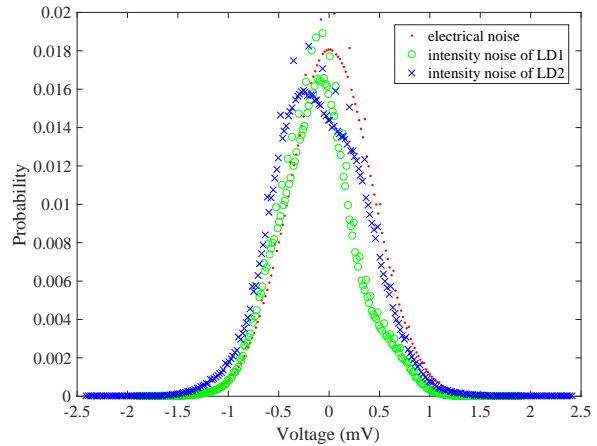}}
\caption{\label{fig:noise}(Color online) \textbf{The observed probability density function (PDF) of the classical noise: electrical noise (red dot), intensity noise of LD1 (green circle), and intensity noise of LD2 (blue cross). }The experimental setups are the same as those of the main text. But we turn off the two lasers for the measurement of the electrical noise, and one of the lasers for the measurement of the intensity noise of LD1 and LD2. The sampling rate of the oscilloscope is 20 GHz, and the recorded sample size is $10^8$.}
\end{figure}

In order to get the bound of the classical noise, we should analysis the types of the classical noise. In fact, there are three types of main classical noise, which are the classical phase noise of interferometer, intensity fluctuation of the light (both LD1 and LD2), and classical electrical noise of the measurement devices (photodetector and oscilloscope).

A interferometer is required to measure the phase of the light, but the mechanical vibration and thermal effect will affect the stability of the interferometer, and introduce classical phase noise. But, in our experiment, such classical phase noise is ignored. That is because the quantum random number is determined by the relative phase of the adjacent sampling point. Thus, we could remove the mechanical vibration and the thermal effect by isolating the interferometer from the environment, as what is generally done in the phase-encoding QKD system. With such methods, the time period is about 3min when the phase of interferometer changes from 0 to p$\pi$. Thus, the difference of the phase between the adjacent sampling point is about $\Delta \theta_c \approx \pi T_s/ 180s$. Here $T_s$ is sampling time. It is easy to get that $\Delta\theta_c$ is much smaller than the random phase coming from the spontaneous emission. For example, in our experiment, $\Delta\theta_c \approx \pi 50ns/180s \approx 2\pi 10^{-10}$ for the CW+CW case, and $\Delta\theta_c \approx \pi 2ns/180s \approx \pi 10^{-11}$ for the CW+Pulse and Pulse+Pulse case. Of course, when the length of the generated bit string is very long, such as lager than 1Tbit, the classical phase noise of the interferometer will introduce classical correlation between the first bits and end bits. Then such classical phase noise must be taken into account in the estimation of min-entropy. However, in our proof-of-principle experiment, such effect is not considered. In other words, the classical phase noise introduced by the interferometer is ignored in our analysis.

The intensity fluctuation of the light will also affect the randomness of the generated bit string. The intensity fluctuation comes from both form the fluctuation of the driven current of laser and the fluctuation of the optical setups (such as fiber, beam splitter, and so on). Furthermore, the classical electrical noise of the measurement devices (photodetector and oscilloscope) will also affect the measured voltage and the output of analogy-to-digital converter (ADC). Thus, in this paper, we only consider the two main classical noise, since they can be directly measured in experiment. Fig.~\ref{fig:noise} shows the PDF of the measured classical electrical noise and intensity noise of lasers in our experiment. It clearly shows that the classical noises are much lower than the amplitude of the quantum signal.

With the experimental data of Fig.\ref{fig:noise}, we could directly get the bound of the classical noise. In fact, with the confidence 99.9999\%, the bounds are $-2.29mV \leq n_e \leq 2.29mV$, $-1.88mV \leq n_{LD1} \leq 1.88mV$, and $-2.07mV \leq n_{LD2} \leq 2.04mV$, for the electrical noise ($n_e$), intensity noise of LD1 ($n_{LD1}$), and intensity noise of LD2 ($n_{LD2}$), respectively. Thus, if we assume the three types of noise (electrical noise, intensity noise) are independent, the bound of the total classical noise can be obtained, which are $n_{min}=-6.24mV$ and $n_{max}=6.21mV$.

According to the analysis given above, if we know the PDF of the quantum signal, we could estimate the min-entropy of genuine quantum randomness. In main text, based on our experimental results, we adopt this model to calculate the min-entropy of genuine quantum randomness for three QRNG cases, cw+cw, cw+pulse and pulse+pulse, respectively.

\begin{figure}
\centering \subfigure[] {\includegraphics [width=7.5cm]
{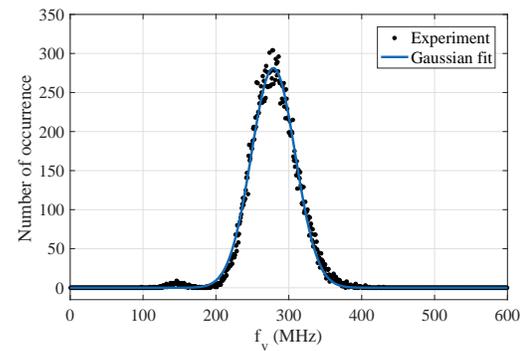} \label{fig:cw2_fv}} \qquad \subfigure[]
{\includegraphics [width=7.5cm] {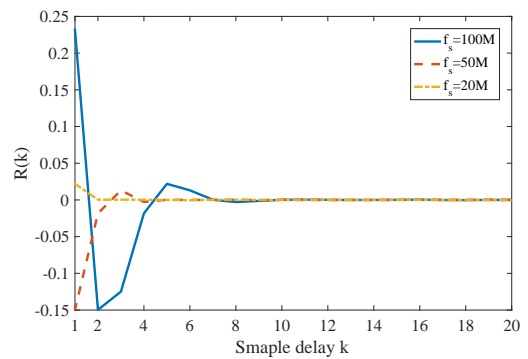}
\label{fig:autoco_diff_fs}}
\caption{ (Color online) \textbf{Measurement results for CW+CW case. }(a) Beating signal's frequency. The sampling rate is 10 GHz in this measurement. The solid line is the Gaussian fitting curve for the experimental data. The beating signal has a mean 278.7 MHz and a standard deviation 30.2 MHz. (b) The autocorrelation coefficient of PD's output with sampling frequencies at 100 MHz (blue solid), 50 MHz (red dash), and 20 MHz (yellow dash-dotted). }
\end{figure}

\begin{figure}
\scalebox{1}{\includegraphics[width=\columnwidth]{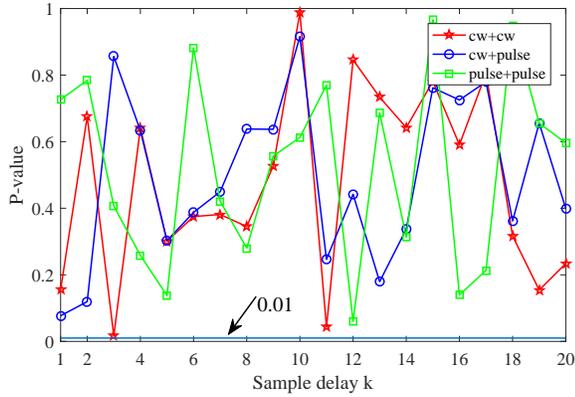}}
\caption{\label{fig:auto_pvalue}(Color online) The P-Value for the autocorrelation coefficient of the extracted bit string. In the calculation, the length of data is $10^8$, and the statistical significance is set as $\alpha=0.01$. All of them are larger than the statistical significance, which means there is insufficient evidence to conclude that there is a significant relationship between the bit string $X=[x_i]$ and $X'=[x_{i+k}]$.}
\end{figure}



\begin{thebibliography}{100}
\bibitem{NIST} In NIST tests, 1000-trails, each with 1 Mbits sample, are tested. Using the significance level $\alpha=0.01$, the P-value should be larger than 0.0001, and the proportion should be lager than 98.06\%. If multiple probability of passing and P-Values are generated in the tests, the worst cases are given. The X-axis labels from 1 to 15 are test terms: Frequency, BlockFrequency, CumultiveSums, Runs, LongestRun, Rank, FFT, NonOverlappingTemplate, OverlappingTemplate, Universal, ApproximateEntropy, RandomExcurisions, RandomExcursionsVariant, Serial, and LinearComplexity.
\end{thebibliography}
\end{document}